# Achilles, Neural Network to Predict the Gold Vs US Dollar Integration with Trading Bot for Automatic Trading

**Angel Varela1\***　　　**Colegio Maria Cano I.E.D.1\***　　　angel.daniel.varela25@gmail.com

| **Keywords** | **Abstract** |
|---|---|
| Machine Learning<br>Neural Network<br>Long Short Term Memory<br>Commodity<br>Gold vs US Dollar<br>FinBERT<br>Financial News<br>Trading Bot | Predicting the stock market is a big challenge for the machine learning world. It is known how difficult it is to have accurate and consistent predictions with ML models. Some architectures are able to capture the movement of stocks but almost never are able to be launched to the production world. We present Achilles, with a classical architecture of LSTM(Long Short Term Memory) neural network this model is able to predict the Gold vs USD commodity. With the predictions minute-per-minute of this model we implemented a trading bot to run during a month of testing excluding weekends. At the end of the testing period we generated $1623.52 in profit with the methodology used. The results of our method demonstrate Machine Learning can successfully be implemented to predict the Gold vs USD commodity. |

## 1. INTRODUCTION

Trying to predict assets is always a trending affair in the machine learning world. There have been many models and methodologies trying to predict the stock or other markets with less or higher accuracy. Even humans cannot certainly know what will be the movement of the market in Cryptos like Bitcoin, Forex like Australian Dollar vs USD or Stocks like Amazon. For these reasons machine learning has become extremely popular for this task. Unfortunately many of the state-of-the-art architectures have had a regular performance. Our challenge is to innovate a field that has been broadly studied by both academia and industry. We want to propose a methodology predicting a commodity in this manner to finally conclude that the machine learning models can predict the market volatility and integrated with a trading bot can earn you a ton of money if used in the right hands.

Gold vs USD is a relatively uncommon market when trying to use machine learning architectures. We noticed this commodity can be more easily understood by our LSTM Model(Achilles). Gold is a limited resource and a large history of data, it can repeat patterns our models can follow and apply forward predictions more easily. Some of the markets that are more difficult to implement are those which do not have any limit. Like Cryptocurrencies(Bitcoin, Ethereum) and Stocks.

Right now there are many stock prediction models out there like ARIMA (*Peiqi Liu(2024), Time Series Forecasting Based on ARIMA and LSTM. Department of Business Management, United International College. Zhuhai, vol 656, pp. 1203-1207*) or Prophet (*Emir Žunić1, Kemal Korjenić , Kerim Hodžić and Dženana Đonko, International Journal of Computer Science & Information Technology, vol 12, no 2, pp. 23-34* ). They focus on single day predictions for only one stock at a time.

We're focusing on predicting each minute of the commodity. We performed a model who's able to output the price each minute the market is open during one month. We will integrate these predictions with a trading bot(Section 4) in an infinite loop running minute per minute to buy when the market is bullish and sell when it is bearish.

## 2. LITERATURE REVIEW

Trying to predict the stock market has become a popular task in the last years, some aspects of these are obviously the opportunity to invest and take profit of desired stocks without knowing too much about the fluctuation of these markets, Saber Talazadeh and Peraković used Random Forest and Sentiment Analysis with FinGPT to enhance the predictions of the stock market, we implemented the sentiment analysis component in our Trading Bot (Section 4.1). Also, Alamu and Kamrul Siam used models including LSTM, GRU, ARIMA, and ARMA to predict stock prices over several time periods. After a detailed experiment they were able to demonstrate LSTM outperform traditional techniques due to their capability to understand the non-linear patterns of the stock market. We are using the LSTM architecture for this reason. Most of the approaches try to measure if the desired stock will go up or down, The models are trained in daily data and are used in forward predictions in the same manner. These models are known for their inaccurate predictions in production, for these reasons this task has been seen as impossible for many academics. Most of the research papers in this area focus mostly on how accurate the models are but are almost never launched into production, We focus on this. We decided to launch this model in a month-long period of testing with a paper account taking into account all the challenges of this task. This paper will demonstrate how LSTM can successfully make accurate predictions for a specific Commodity, Gold vs US Dollar, we will integrate two components, the predictions of Achilles and the news from FinBERT into a trading bot.

**2.1 Jim Simons and MEDALLION FUND**

Jim Simons was a mathematician and the chairman of the Medallion fund. One of the most profitable and bigger hedge funds in history. The fund started in 1988 and since then they were able to grow to a cap of $114 billion in 2018. The fund used ML and other technologies such as quantitative analysis to make automated systems for algorithmic trading.

The fund started its formal operations more than 30 years ago, their investment strategy is still unknown. Many secrets are around their technology and strategy. In [25] the author describes how his ideas with ML and Algorithmic trading reshaped the world of trading. This paper is inspired by their work and how, 30 years ago, they were able to have such an important impact in the world of finance.

**3. METHODOLOGY**

We present our approach in this section, note that our model is trained on time series data minute per minute and explained in the Section 4.3 investing in the MetaTrader Terminal*(MetaTrader5 (2024, 10, 8) Trading Terminal https://www.metatrader5.com/en )*

**3.1 Data**

The market we chose to train our model is ETH-USD (Ethereum), known by its volatility. The reason we chose this currency is the large amount of data and also the unpredictability of its nature. Cryptocurrency are so volatile, the cost of only one Ethereum on march, 2016 was $10.04, at the date of this paper is $2,580.72.

We used data from 3 different timeframes. 15-Minutes Data, 5-Minutes data and 1-Minute data. The precise structure of our data is shown in Table 1(Timeframe minute per minute). Table 2 has the number of rows per each of these Timeframes.

**Table 1. Data used to train our model (Timeframe minute per minute)**

| Time | Open | High | Low | Close | Volume | RSI | EMA |
|---|---|---|---|---|---|---|---|
| 2024-06-01 4:51:00 | 3757.48 | 3756.14 | 3758.68 | 3758.68 | 143 | 52.606645 | 758.393539 |
| 2024-06-01 4:52:00 | 3758.34 | 3757.12 | 3759.93 | 3759.93 | 151 | 58.774908 | 3758.527144 |
| 2024-06-01 4:53:00 | 3760.29 | 3758.59 | 3759.74 | 3759.74 | 112 | 57.475447 | 3758.63261 |
| 2024-06-01 4:54:00 | 3759.74 | 3758.8 | 3759.53 | 3759.53 | 76 | 55.913922 | 3758.710644 |
| 2024-06-01 4:55:00 | 3759.93 | 3758.75 | 3759.87 | 3759.87 | 74 | 58.024096 | 3758.811458 |
| 2024-06-01 4:56:00 | 3759.87 | 3758.05 | 3760.31 | 3760.31 | 74 | 60.799189 | 3758.941766 |
| 2024-06-01 4:57:00 | 3760.42 | 3759.29 | 3760.33 | 3760.33 | 136 | 60.933358 | 3759.062482 |
| 2024-06-01 4:58:00 | 3760.8 | 3758.96 | 3761.38 | 3761.38 | 148 | 67.589146 | 3759.264005 |
| 2024-06-01 4:59:00 | 3761.28 | 3760.43 | 3761.97 | 3761.97 | 110 | 70.785445 | 3759.499309 |
| 2024-06-01 5:00:00 | 3761.88 | 3761.16 | 3764.07 | 3764.07 | 188 | 79.149716 | 3759.89676 |

We are using the open, high, low, close, volume, RSI and EMA for training, the last two are used as technical indicators(Section 3.2) and the other 5 will help our model make more accurate predictions not only on the test dataframe but on the validation data

**Table 2. Data distribution**

| TimeFrames | # Rows |
|---|---|
| ETH-USD 15-Minutes Time Frame | 125160 |
| ETH-USD 5-Minutes Timeframe | 125160 |
| ETH-USD 1-Minute Timeframe | 432115 |
| Total | 682434 |

A total of 682.434 rows are used to train Achilles. We decided to use data from January 2018 in the 15 minutes dataframe and 5 minutes dataframe and minute-per minute data from January of 2024. **3.2 Technical Indicators**

We are using the RSI(Relative Strength Index) and EMA (Exponential Moving Average) as technical indicators to enhance the predictions of our model. We are using different Time Frames because we want our model to learn the seasonality of the data.

### *3.2.1 RSI(Relative Strength Index)*

To calculate the RSI for each trading period an upward change *U* or downward change *D* is calculated. Up periods are characterized by the close being higher than the previous close. The Upward periods are calculated as calculated as $U = \text{close}_{now} - \text{close}_{previous}$ and Downward periods are $D = \text{close}_{previous} - \text{close}_{now}$. Using an n-period SMMA(Smoothed or Modified Moving Average). The ratio of these averages is the relative strength or Relative Strength Factor:

$$RS = \frac{\text{SMMA}(U,n)}{\text{SMMA}(D,n)}$$

The relative strength factor is then converted to a Relative Strength Index between 0 and 100

$$RSI = 100 \cdot \frac{\text{SMMA}(U,n)}{\text{SMMA}(U,n) + \text{SMMA}(D,n)} = 100 - \frac{100}{1 + RS}$$

### *3.2.2 EMA(Exponential Moving Average)*

The Exponential Moving Average is great when trying to predict volatility in Cryptocurrencies and others such as Stocks and Forex. We will use it to improve the predictions of our model

$$EMA = Price(t) \times k + EMA(y) \times (1 - k)$$

Where *t* stands for today, *y* for tomorrow, $k = 2 \div (X - 1)$ and *X* is the number of days in EMA.

### 3.3 Data Preprocessing

To preprocess our data we're using a sliding window approach, this consists of sliding our data into smaller subsequences to feed into our LSTM Model (Long Short Term Memory). We forecast the next period of the input data with this formula:

*Xi={dataset[i+j,:]|j=0,1,...,N−1}*
*Yi=dataset[i+N,−1]*

Where *N = 120, (i)* is each element of the dataset. This approach will help to have more accurate predictions. We use *N = 120* to use the last 120 minutes to train on the next minute.

### 3.4 Model (Achilles)

Achilles is a simple model with three layers, an input layer, a hidden layer and a Dense Output Layer, Figure 1 shows the architecture of our model, table two it's results predicting the test dataset in ethereum, we found this model to have the best performance after predicting Commodities such as Gold vs USD, ETF such as S&P500 and Forex like USD vs Hong Kong Dollar with the highest accuracy. Figure 2 demonstrates the predictions of our model in Ethereum over a period of two months. Note that our model only has 9.544 parameters, it is a small model, this will make it easier when training and predicting the future days for computational resources.

**Figure 1. (Architecture of our model)**

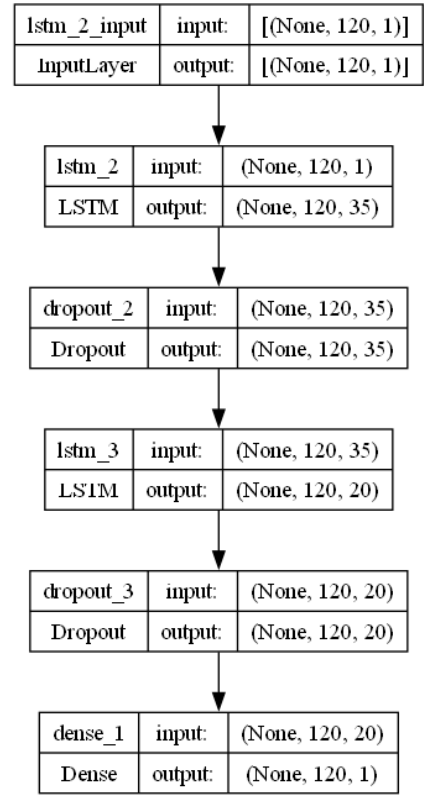

Our model consist of a input layer with input shape (None, 120, 35), it has 35 neurons in the first layer, 20 neurons in the hidden layer and 1 dense output neuron on the output layer, we found this structure to have the fastest computation and higher accuracy all among other architectures.

**Figure 2 (Predictions in test dataset of ETH-USD)**

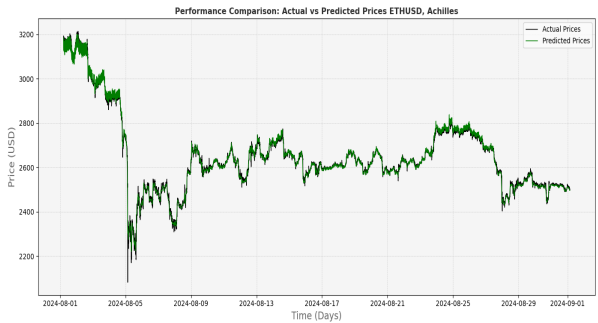

Figure 2 demonstrates the predictions of our model in the rest dataset during 2 months, the estimations are almost perfect on the test dataset. Note is tested on the same training symbol ETH-USD.

### 3.5 Predictions

We are using a minute-per-minute approach. Our model will be predicting each minute and based on these predictions we'll let our Trading Bot decide if to invest or not. Table 3 shows what our predictions look like.

**Table 3. (Predictions on the future of Achilles)**

| Date | Price |
|---|---|
| 2024-08-26 0:00:04 | 2318.995605 |
| 2024-08-26 0:01:04 | 2317.996094 |
| 2024-08-26 0:02:04 | 2317.996094 |
| 2024-08-26 0:03:04 | 2318.488281 |
| 2024-08-26 0:04 | 2317.027588 |
| 2024-08-26 0:05:04 | 2317.630371 |
| 2024-08-26 0:06:04 | 2315.318359 |
| 2024-08-26 0:07:04 | 2314.81543 |
| 2024-08-26 0:08:04 | 2314.857422 |

The predictions you can see above are extremely important for the trading bot as it will determine when we'll invest or not. In order to predict two months of one stock we need to predict 33.000 minutes. The predictions we use are stacked, we'll predict the first minute based on the last 120 minutes of the passed dataframe, and add the minute we predicted to predict the second minute and then on, until the point we reach 33.000, with this approach our model is able to accurately predict into the future the price of the Gold vs US Dollar.

### 3.6 Sentiment Analysis with FinBERT

In addition we're implementing FinBERT who's a model trained specifically to estimate the sentiment from financial news. We're parsing news from Three different websites.

1.Benzinga.com, 2.Investing.com, 3.Ft.com. All three websites are recognized for being unbiased and have real-time news. In Section 3.7 we'll retrieve and estimate the sentiment of these three financial news websites each 15 minutes.

The data consist of Probability and Sentiment respectively, we're using a simple formula to calculate the average probability of all three websites.

$$A = \frac{1}{n}\sum_{i=1}^{n} a_i$$

$A$, is the Arithmetic mean, $n$ is the number of probabilities and $a_i$ are the probabilities. Table 4 shows our probabilities taken each 15 minutes.

**Table 4. (Probabilities in 15 Minutes Dataframes)**

| Time | Average Probability | Probability Benzinga.com | Probability Investing.com | Probability Ft.com |
|---|---|---|---|---|
| 2024-10-02 21:02:04 | 0.8754457653 | 0.8615154028 | 0.9618232846 | 0.8922206759 |
| 2024-10-02 21:14:04 | 0.8754324138 | 0.8615154028 | 0.9617832303 | 0.8922206759 |
| 2024-10-02 21:35:04 | 0.8754324138 | 0.8615154028 | 0.9617832303 | 0.8922206759 |

The sentiment is evaluated as follows. *Positive = 1, Negative = -1, Neutral = 0.* Then we convert each sentiment into this numerical value and use the same formula above to extract the Average Sentiment, We will implement this procedure in Section 4.1.2.

### 4. EXPERIMENTS (Trading Bot)

We merge all together in our Trading bot, we connect to the TradingTerminal(We will not go all over the connection procedure with the terminal for explanation purposes) and run two infinite loops each minute(Section 4.2) and each 15 minutes(Section 4.1), It is important to keep in mind that we're using a paper account for the experiments and we do not have the fees and commissions real accounts could have.

### 4.1 Inner Loop

The Inner loop is used to extract the data from different websites*(Benzinga.com. (2024, 10, 8). XAUUSD. https://www.benzinga.com/, Investing .com. (2024, 10, 8). XAU-USD. https://www.investing.com/, ft.com. (2024, 10, 8). US Dollar. https://www.ft.com/)* and feed into FinBERT(Section 3.6), We implement the headlines and contents from market analysts and indicators. See Table 4

Our Inner Loop will run each 15 minutes since we want our trading bot to have the latest news of the Commodity. The data is then parsed in text format and we extract the Probability and Sentiment. Probability is a percentage of the probability that the market will go up or down, if the probability is higher that means the market will go up, if it's lower it means it will go down. We are using the average and overall sentiment explained in section 3.6 when sending orders(Section 4.2.3)

We decide to use the sentiment to determine if we want to buy or sell in our infinite interaction. If the Sentiment is positive we will buy, if it's negative we'll sell(Section 4.2.3)

**4.2 Outer loop**

This loop will run each minute and is where everything is working. We'll Extract the data(Section 4.2.1), use Positional Sizing(Section 4.2.2) and send orders(Section 4.2.3) Below you will find the detailed explanation of our Trading Bot, Outer loop running minute-per minute and the Inner Loop Running each 15 Minutes.

**Figure 3. Architecture of our Trading Bot**

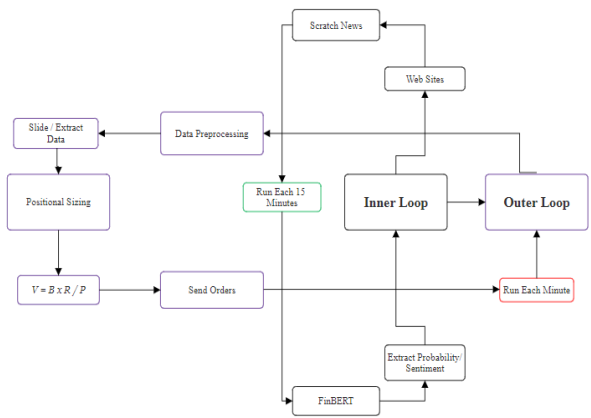

*4.2.1 (Extracting Predictions Data)*
We use the data(Table 5) and slide it in rows of each 20 minutes, 10 last minutes based on the now time and 10 next minutes based on the now's time.

Note that we take a look at this challenge as a Regression problem, We want to get the price minute per minute, in a classification procedure we would ask our model to know if the price will go up or down. we're retrieving the real value in the next minutes. We extract the lowest and max price of this slide, we'll need this for Section 4.3

**Table 5. (Predictions of Nearest minutes from now)**

| Data | Price |
|---|---|
| 2024-09-25 21:39:04 | 2437.80664 |
| 2024-09-25 21:40:04 | 2436.84375 |
| 2024-09-25 21:41:04 | 2436.603271 |
| 2024-09-25 21:42:04 | 2436.124267 |
| 2024-09-25 21:43:04 | 2435.044921 |
| 2024-09-25 21:44:04 | 2435.39746 |
| 2024-09-25 21:45:04 | 2436.124267 |
| 2024-09-25 21:46:04 | 2436.603271 |
| 2024-09-25 21:47:04 | 2436.478027 |
| 2024-09-25 21:48:04 | 2436.124267 |
| 2024-09-25 21:49:04 | 2435.285644 |
| 2024-09-25 21:50:04 | 2435.163574 |
| 2024-09-25 21:51:04 | 2434.686523 |
| 2024-09-25 21:52:04 | 2434.558593 |

Conditional if $x = a$ maximum of $f(x)$
$$f(x) \leq f(a) \text{ for all } x \in D$$

$f(x)$ is the maximum value of the function
Conditional $x = a$ if minimum of $f(x)$
$$f(x) \geq f(a) \text{ for all } x \in D$$

$f(x)$ is the minimum value of the function, we later use the minimum and maximum to buy and sell respectively, we decided to apply this same formula from the real price dataframe we take from MetaTrader 5. We will use the last 17 minutes for two reasons.

First to take the latest price, it needs to be used to measured the quantity or volume we'll use to invest in the infinite interaction, second to check if the current real minute has the lowest or highest price, this procedure is explained in Section 4.2.3

*4.2.2 Positional Sizing*

To calculate the volume we will trade we use a simple formula, with a risk we can set up to 1.0 or 100%

$$V = \frac{B \times R}{P}$$

Where $V$ is the volume, $B$ is the balance of the account, in this experiment is set to 1000, $R$ is the risk(The percentage of balance we'll use to invest), set to 0.3 and $P$ is the current price of the Commodity.

### *4.2.3 Send Orders*

We send orders Through MetaTrader 5 in a paper account. The logic we use to invest is implemented as follows. We use the Inner Loop to see the probability and sentiment. If the sentiment is positive and the probability is higher than 0.87 we may buy, if the sentiment is negative and the probability is lower than 0.5 we may sell. Otherwise we keep the loop running.

When there are open positions with a buy ticket we close them when the now's price is the highest in our prediction dataframe, if the real price is the lowest then we do not close them.

Otherwise, a sell position is closed when the now's price is the lowest of the slice in the predicted dataframe. If the real price is the highest we don't close the order. We implement these two procedures when closing the two orders with highest profit in the paper account. Look at Figure 4. Note we use the volume on Section 4.2.2

**Figure 4. FlowChart when sending an order**

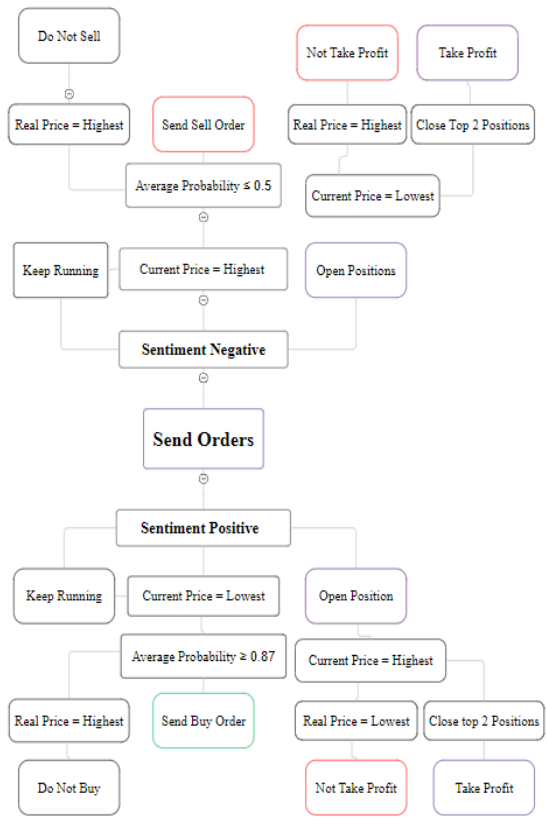

### 4.3 Overview

We use the predictions from Achilles along with the sentiment and probability from Finbert into our trading bot. We extract the min and max minute from a slide of 20 minutes based on the current minute. Then if the triggers are activated, See Figure 4. We Buy, Sell and Close the orders. We implemented this procedure in the Gold Vs USD Market as it's a limited Commodity is relatively easy to predict. We Extract the Probability and Sentiment From three websites each 15 minutes and keep running our trading bot each minute during one month excluding weekdays. From September 1st, 2024 to September 30th, 2024.

## 5. RESULTS

In the test period we finished with a total of $1623.52 in profit in one month. We use an initial budget of $1000 and a risk of 0.3, the results on Figure 5 shows the profit over time

In our experiment, the Trading Bot has a big impact on these results since the orders need to be closed in an appropriate manner.

**Figure 5 (Chart of profits)**

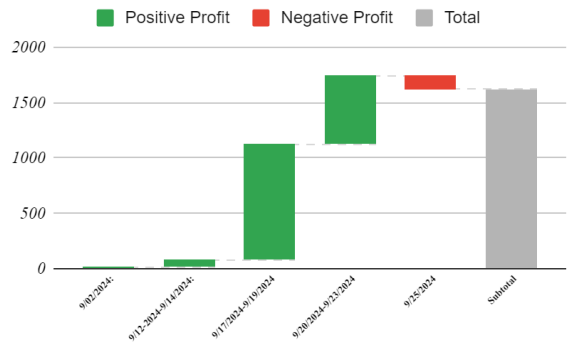

The waterfall chart shows the great performance of our trading bot with the predictions of Achilles. Note that the risk is pretty high, we're using 30% Of our balance in each one of our trades. From September 2nd to 14th the profit was relatively low, in the period of September 17th to 19th the profit went considerably up due to many orders closed correctly by our model(Figure 6). In September 20th to 23th we also did have a good performance and we closed our results with a negative profit on September 25th, the last 5 days the bot determined it was not worth it to trade.

### 5.1 Explanation

Our model is trained to do shorting. As you have noticed we have the minute per minute predictions because we want to buy when the price is low in the chart and then automatically close the orders when the price is higher, this procedure can be easily demonstrated in Figure 6.

**Figure 6 (Chart of profits on August 17th to 19th 2024)**

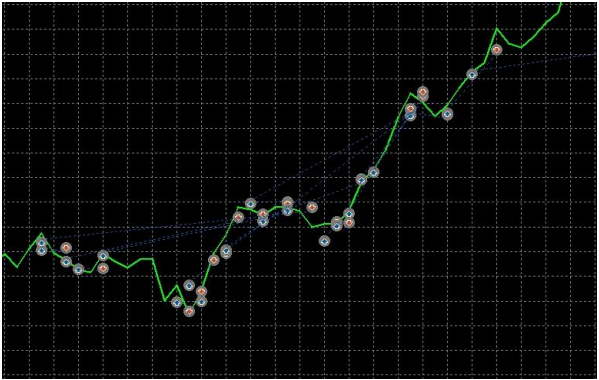

Figure 6 shows the performance of our model in the timeframe of 15 minutes. Our model has enough patience to close the orders when the price goes up. When the price is low it does not close the orders. This demonstrates it has enough patience to keep an order running for days and even weeks and close them at the right moment. The Trading bot invests in a manner where if the market is bearish we sell, if it's bullish we buy. Look at Figure 7.

**Figure 7.(TimeFrame 15 minutes, Patience of our trading bot)**

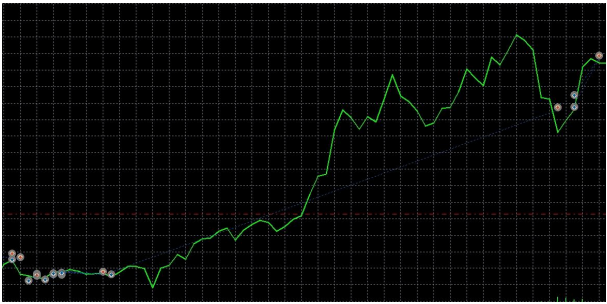

The Trading bot is also set to close the first 2 orders with the highest profit, even if in the predicted data frame the price is the higher, if the real price is the lower the trading bot will not close the orders. By doing this we ensure the trades will not be automatically closed even if the price on the predicted dataframe is the lowest. One of the main challenges of investing is knowing when to close your orders. Our trading bot successfully applied these principles in Figure 8th.

**Figure 8.(Orders placed on September 20th to 23rd)**

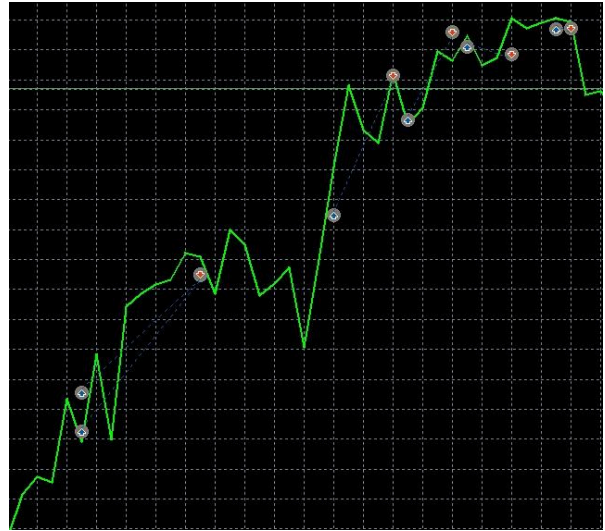

It demonstrates it's effectiveness of LSTM when predicting the Gold vs USD market, in production our method is able to make a ton of money with almost no intervention. We define our perfect trade when the bot buys when the price is low and closes the order when the price has gone up just right after, this behavior can be demonstrated in Figure 9.

**Figure 9. (Order sent in September 18th)**

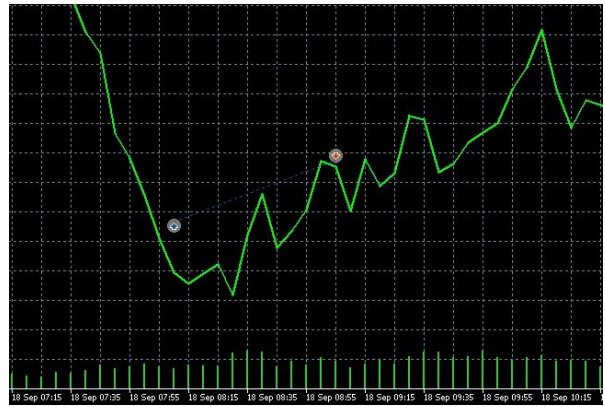

As you can see the Trading Bot is able to open the order when the prices are low and then take the profit. The trading bot is characterized for doing these orders in short periods of time, that way it makes a high profit in a reduced amount of days when determinants so with the FinBERT Sentiment. Even if our method demonstrates its effectiveness in production, it does not mean we have losses in our procedure. Over 70% of people lose money in trading. The method is able to make a high profit in the shorting strategy but also is able to lose almost all the balance in a short term period. Look at Figure 10.

**Figure 10. (Profit in September 25th)**

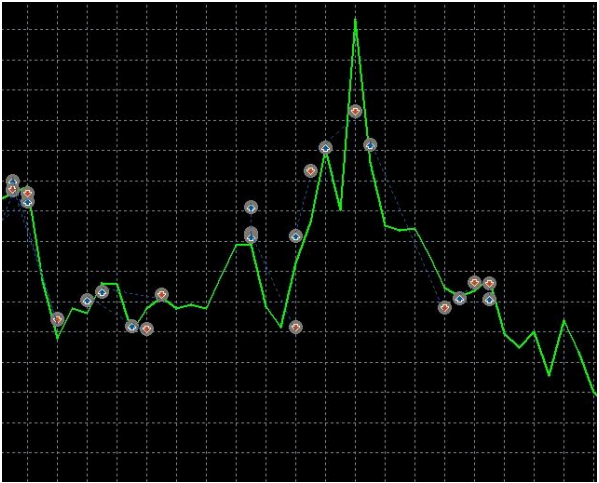

With these results we can conclude that LSTM models can be integrated into trading bots to send automated orders without any human intervention and still have reliable and accurate performance, based on our balance we were able to take 162% of profit with this strategy in just one month.

**5.2 Overview**

Our trading bot with the predicted dataframe from Achilles was able to outperform some of the latest models in this field. We used a minute-per-minute approach for shorting, we used the sentiment of three different websites to estimate if the market is bullish or bearish and make orders based on this. If the market is neutral we do not place any order. Lastly the procedure did have a bad performance in the last days of the experiment with a total loss of $119.57.

**Table 6. (Profits in our test period)**

| Dates | Profit |
|---|---|
| 9/02/2024: | 19.27 |
| 9/12-2024-9/14/2024: | 66.31 |
| 9/17/2024-9/19/2024 | 1044.16 |
| 9/20/2024-9/23/2024 | 611.35 |
| 9/25/2024 | -119.57 |
| **TOTAL** | **1623.52** |

In summary our model did have a consistent performance in the paper account taking 1623.52 in profit at the end testing period, we started with a reasonable income from September 2nd to September 14th, from September 17th to 23rd the bot generated $1655.51, on September 25th we finished with losses. The rest of the days the bot determined was not the moment to invest.

Table 7 determines the performance of our model in comparison to other models.

**Table 7. (Performance of our model in comparison to other approaches)**

| Approach | Testing loss | MAE | MAPE |
|---|---|---|---|
| Fin-BERT Embedding LSTM | 0.00083 | 173.67 | 0.045 |
| LSTM | 0.00092 | 183.36 | 0.072 |
| DNN | 21.77 | 489.32 | 0.22 |
| Achilles | 0.0033 | **22.905** | **0.009** |

Our model outperformed some of the latest models with news integrated, the Mean Absolute error and Mean Absolute percentage error surpasses other models performance. We demonstrated this approach has even better results in production in a paper account over the testing period with quite impressive results.

**6. CONCLUSION**

In this paper we've concluded Achilles has an accurate performance when predicting the Gold Vs USD Commodity market.

We started with a budget of $1000 with a risk of 0.3 and finished the test period with $1,623.52 in profit. We integrated a trading bot who's able to close the orders just in the right moment to take the profit. Note that this model is only used on the Gold Vs USD Commodity, and has not been tested on Forex or Stocks. Achilles only has 9.544 parameters and is not trained in a huge dataset, in addition it is important to keep in mind that this is a paper account and we do not encourage trading real money as Section 5.1 demonstrates the inconsistency of the LSTM model can lead to poor results. The implications of this model can be quite notorious in the trading word. If LSTM models without too many parameters and relatively low datasets are able to predict some markets it could impact the way we see trading. Further results need to be used to demonstrate the effectiveness of LSTM models to predict Commodities. This paper demonstrates the predictions of LSTM integrated with Trading Bots can generate profitable results in the short term in the Gold vs US Dollar(XAUUSD).

**7. REFERENCES**

[1]     Peiqi Liu (2024), Time Series Forecasting Based on ARIMA and LSTM. Department of Business

Management, United International College. Zhuhai, vol 656, pp. 1203-1207

**[2]** Emir Žunić1, Kemal Korjenić , Kerim Hodžić and Dženana Đonko (2020), International Journal of Computer Science & Information Technology, vol 12, no 2, pp. 23-34

**[3]** Deri Siswara , Agus M. Soleh , Aji Hamim Wigena (2024), Classification Modeling with RNN-based, Random Forest, and XGBoost for Imbalanced Data: A Case of Early Crash Detection in ASEAN-5 Stock Markets , International Journal of Computer Science & Information Technology, Vol 12, No 2, pp. 569-579

**[4]** Kelvin J.L. Koa, Yunshan Ma∗, Ritchie Ng, Tat-Seng Chua(2023), Diffusion Variational Autoencoder for Tackling Stochasticity in Multi-Step Regression Stock Price Prediction, National University of Singapore, Eastspring Investments, National University of Singapore, pp. 1-15

**[5]** Wenjun Gu, Yihao Zhong, Shizun Li, Changsong Wei, Liting Dong, Zhuoyue Wang, Chao Yan (2024), Predicting Stock Prices with FinBERT-LSTM: Integrating News Sentiment Analysis, Department of Electrical and Computer Engineering, Northeastern University, pp. 1-9

**[6]** Xiaolin Hu and P (2008). Recurrent Neural Networks . Intechopen

**[7]** Xinyi Li , Yinchuan Li , Hongyang Yang , Liuqing Yang , Xiao-Yang Liu(2019), DP-LSTM: Differential Privacy-inspired LSTM for Stock Prediction Using Financial News ,Columbia University, Beijing Institute of Technology, pp. 1-8

**[8]** Sidra Mehtab , Jaydip Sen and Abhishek Dutta(2020), Stock Price Prediction Using Machine Learning and LSTM-Based Deep Learning Models , Department of Data Science and Artificial Intelligence, Praxis Business School, pp, 1-16

**[9]** Ivan Letteri(2023), VolTS: A Volatility-based Trading System to forecast Stock Markets Trend using Statistics and Machine Learning, University of L'Aquilam, pp. 1-10

**[10]** Yue-Gang Songa , Yu-Long Zhoub∗ , Ren-Jie Hanc(2018), Neural networks for stock price prediction, Business School, Henan Normal University, School of Mathematics, Yunnan Normal University, College of Economics, Sichuan University, pp. 1-11

**[11]** Xiaobin Tang, Nuo Lei, Manru Dong, and Dan Ma(2022), Stock Price Prediction Based on Natural Language Processing, School of Statistics, University of International Business and Economics, vol 2022, pp. 1-1

**[12]** Chaher Alzaman(2023), Forecasting and optimization stock predictions: Varying asset profile, time window, and hyperparameter factors, Department of Supply Chain and Business Technology Management, Soft computing letters, pp. 1-16

**[13]** Andriy Burkov (2019), The hundred Page Machine Learning Book

**[14]** Alamu, O.S. and Siam, M.K. (2024) Stock Price Prediction and Traditional Models: An Approach to Achieve Short-, Medium- and Long-Term Goals. Journal of Intelligent Learning Systems and Applications, vol. 16, 363-383. https://doi.org/10.4236/jilsa.2024.164018

**[15]** Saber Talazaded and Dragan Peraković(2024), SARF: Enhancing Stock Market Prediction with Sentiment-Augmented Random Forest, British Columbia Institute of Technology, University of Zagreb, pp 1-10

**[16]** Nicholas Renotte, (Jan 19, 2024), How to code an AI trading Bot, Youtube. https://www.youtube.com/watch?v=c9OjEThuJjY&list=WL&index=12&t=1861s

**[17]** Neural Nine, (Oct, 28, 2022), Predicting Stock Prices with FBProphet in Python, Youtube. https://www.youtube.com/watch?v=03H2_ekdv2I&

**[18]** Computer Science, (May, 23, 2023) Use Artificial Intelligence (AI) to Predict the Stock Market with Python https://www.youtube.com/watch?v=fGLY0dIHJ2w&list=WL&index=6&t=469s

**[19]** KothaEd, (Oct 13, 2023) Machine Learning Project in Python to Predict Stock Price, Youtube, https://www.youtube.com/watch?v=P3JlMWoP3fE&list=WL&index=7&t=2411s

**[20]** DataQuest, (May 23, 2022), Predict The Stock Market With Machine Learning And Python, Youtube, https://www.youtube.com/watch?v=1O_BenficgE&list=WL&index=5

**[21]** Benzinga.com. (2024, 10, 8). XAUUSD. https://www.benzinga.com/

**[22]** Investing .com. (2024, 10, 8). XAU-USD. https://www.investing.com/

**[23]** ft.com. (2024, 10, 8). US Dollar. https://www.ft.com/

**[24]** MetaTrader5 (2024, 10, 8) Trading Terminal https://www.metatrader5.com/en

**[25]** Gregory Suckerman, (2019) The Man Who Solved the Market: How Jim Simons Launched the Quant Revolution, Portfolio Penguin.